\documentclass[twocolumn,aps,superscriptaddress]{revtex4-2}
\usepackage[T1]{fontenc}
\usepackage[utf8]{inputenc}
\usepackage[english]{babel}
\usepackage{amsmath}
\usepackage{amssymb}
\usepackage[export]{adjustbox}
\usepackage{bm}
\usepackage{braket}
\usepackage{xcolor}
\usepackage{ytableau}
\usepackage[colorlinks=true,urlcolor=blue,citecolor=blue,linkcolor=blue]{hyperref}

\renewcommand{\i}{\mathrm{i}}

\newcommand{\freiburg}{Physikalisches Institut, Albert-Ludwigs-Universit\"at Freiburg,\\ Hermann-Herder-Stra\ss e 3, D-79104 Freiburg, Germany}
\newcommand{\eucor}{EUCOR Centre for Quantum Science and Quantum Computing, Albert-Ludwigs-Universit\"at Freiburg,\\ Hermann-Herder-Stra\ss e 3, D-79104 Freiburg,  Germany}

\begin{document}

\title{Thermodynamic state variables from a minimal set of quantum constituents}
	
	\author{Uwe Holm}\affiliation{\freiburg}
	\author{Hans-Peter Weber}\affiliation{\freiburg}
	\author{Morgan Berkane}\affiliation{\freiburg}\affiliation{\eucor}
	\author{Camilla Wulf}\affiliation{\freiburg}
	\author{Anton Kantz}\affiliation{\freiburg}
	\author{Anja Kuhnhold}\affiliation{\freiburg}
	\author{Andreas Buchleitner}\email{a.buchleitner@physik.uni-freiburg.de}\affiliation{\freiburg}\affiliation{\eucor}
	
	\date{\today}


\begin{abstract}
We show how the macroscopic state variables pressure, entropy and temperature of equilibrium thermodynamics 
can be consistently derived from the (quantum) chaotic spectral structure 
of one or two particles in 
two-dimensional domains. This provides 
a definition of work and heat from first principles, a microscopic underpinning of the first and second law of thermodynamics, 
and a transparent illustration of the ``eigenstate thermalization hypothesis''.
\end{abstract}

\maketitle



\section{Introduction}
Equilibrium thermodynamics is a remarkably reliable theory -- given its deliberate ignorance of microscopic details of the ensemble dynamics 
it sets out to predict. Built on few basic assumptions such as ``molecular chaos'' (see, e.g., p.~177 in \cite{sommerfeld1988})
and hierarchically organized time scales, it inter alia states, with 
deterministic precision, though using purely probabilistic arguments, that none of the audience in a large and high lecture hall must be afraid that all the 
available air molecules would suddenly pile up within a narrow (and unreachable, thus unbreathable) layer right under the ceiling. Such predictions 
per se impressively underpin the fact that exponentially suppressed event probabilities do imply hard facts, much as deterministic equations do. 

A cornerstone of the textbook derivation of equilibrium thermodynamics is the 
principle of equal a priori probabilities \cite{sommerfeld1988,kittel1984,englert2020}, 
which states that, at any instance of 
time, all microstates of the considered system which are consistent with its macroscopic state are equally likely. On the time axis, this implies some coarse 
graining and the above-mentioned hierarchy of time scales, such as to talk of quasi-static state transformations of the equilibrium state -- which disregard 
the equilibration process
occurring on finite, but sufficiently short time scales, after a perturbation. Microscopically, 
equal a priori probabilities are
thus traditionally conceived as a consequence of the collision-induced, chaotic many-body dynamics of the considered system's -- often a gas -- 
constituents \cite{sommerfeld1988,kittel1984,englert2020}.

Consequently, thermodynamics is in many respects considered as an emergent theory, where coarse graining over the (too) many microscopic degrees of freedom
and associated short time scales is not only unavoidable, but also allows to extract a comparably much smaller number of macroscopic characteristics 
of the multi-constituent system under study -- which yet suffices for its 
robust description and control.

It is therefore
self-evident to ask for the typical scales on which thermodynamic behaviour emerges, and, vice versa, below which the microscopic features of the 
system constituents prevent a sharp definition of state variables, or enforce microscopic -- ultimately quantum -- corrections to thermodynamic predictions \cite{sommerfeld1988}. 
A prominent research direction with dedicated focus on this type of questions is ``quantum thermodynamics'' \cite{gemmer2009}, in essence the study 
of the thermodynamic properties 
of open quantum systems \cite{breuer2002} -- which is systems with at least some of their degrees of freedom coupled (i.e., opened) to some environmental degrees of freedom. 
However, much as thermodynamics itself, these approaches frequently rely on an already established effective description of the open system, by tracing 
over the environment,
and by averaging over (too) short time scales \cite{cohen1996}. In this sense, they do not resolve the microscopic dynamics on the most elementary level. 
Another approach investigates equilibration phenomena (often called ``thermalization'' \cite{rigol2008} -- a terminology which we here 
abstain from) in unitary, interacting multi-constituent quantum systems, as well as deviations thereof (having multiple potential causes) lumped together under the terms ``many-body-localization'' \cite{evers2023}
and/or ``-scars'' \cite{hummel2023,papoular2023,evrard2024,lu2025}.
This research, however, does hitherto not fully establish the connection to standard thermodynamics --  even though attempts to the introduction of certain state variables, such as in \cite{burke2023}, have been made.

\section{Model and Numerical Methods}
Our present contribution establishes the minimal ingredients of a microscopic quantum model which allow for the consistent definition of 
thermodynamic state variables {\em directly} from the spectral and eigenvector structure of one single or two interacting particles confined to 
a two-dimensional -- in the latter case partitioned -- domain. 
We will see that single-particle (quantum) chaos allows for the 
definition 
of pressure, while (interacting) two-particle quantum chaos allows for a microscopic definition of heat flux and, consequently, of entropy and
temperature, ultimately leading to the first and second law of equilibrium thermodynamics.
All spectral information is generated by exact numerical diagonalization of the associated eigenvalue problems, in 
a finite element representation \cite{brugger2024,wulf2024,kantz2025,weber2025,holm2025}.
We set $\hbar \equiv m \equiv 1$,
such that all quantities are given
in terms of the
reference length scale $\mathcal{L}$, set by the geometric extension of the particle's domain. Energy is then
measured
in units of $\mathcal{L}^{-2}$, time in units of $\mathcal{L}^2$, and interaction strength in units of $\mathcal{L}^{-1}$. By additionally setting $k_B\equiv 1$, temperature is also measured in units of $\mathcal{L}^{-2}$.

\section{Isotropy of Pressure and Boyle-Mariotte's Law}
Let us first consider a single particle freely moving in a two-dimensional domain, i.e., a single-particle billiard, with hard walls at which the particle undergoes specular reflection.
Given a rectangular billiard, the particle's classical dynamics is integrable, such that its momentum components along the billiard's walls are constants of motion, giving 
rise to associated good quantum numbers on the quantum level. This cannot lead to a consistent definition of 
pressure \footnote{In contrast to what is implied in many text books, see, e.g., \cite{kittel1984,bartelmann2015}!},
since the generalized force with which the particle acts upon the billiard's walls (and, hence, pressure) is in general not isotropic, as a direct consequence of the 
particle's integrable dynamics (or of the absence of ``molecular chaos'' \cite{sommerfeld1988}) \cite{wulf2024}. 

It is therefore necessary to break the integrability of the motion by a choice of boundary 
conditions which induce ergodic dynamics. Indeed, if a Sinai \cite{sinai1970} instead of a rectangular billiard is quantized, the 
particle's wave vector's direction (its modulus being constant due to
energy conservation) gets randomized by the classical dynamics. On the quantum level, this leads to ergodic eigenstates which are delocalized over the billiard's area,
and can be understood as random superpositions of plane waves \cite{berry1977,mcdonald1988}. 

Consequently, and as an instructive illustration of the eigenstate thermalization hypothesis \cite{deutsch1991,srednicki1994}, the pressure
$P$ imparted by the quantized particle's motion upon the billiard's walls can be derived as follows:
At energy $E=p^2/2$, and for the Hamiltonian 
$H=-\Delta /2$,
with Sinai's boundary conditions,
the standard definition of equilibrium thermodynamics \cite{bartelmann2015} can be rephrased in terms 
of the
Hellmann-Feynman theorem \cite{hellmann1937,feynman1939}, which directly extracts pressure from the energy level's parameter dependence:
\begin{eqnarray}
P & = & - \left\langle \frac{\partial H(p;\lambda)}{\partial \lambda}\right\rangle \nonumber \, , \\
& = &  -  \left\langle E(\lambda)\bigg|\frac{\partial H(p;\lambda)}{\partial \lambda}\bigg|E(\lambda)\right\rangle  \nonumber \, ,\\ 
& = & - \frac{\partial E(\lambda)}{\partial\lambda} \, .
\label{HF}
\end{eqnarray}
In these relations, $E(\lambda)$  is the energy eigenvalue of the Hamiltonian $H(\lambda)$ with associated eigenvector $\ket{E(\lambda)}$,
parametrized by $\lambda$, which in the here considered specific scenario can be the length of either one of the straight walls of the 
Sinai billiard (reduced to its elementary domain, i.e., a rectangle with a quarter-circle removed from one corner) 
or the quarter-circle's radius, allowing to vary the billiard's two-dimensional volume \cite{wulf2024,kantz2025,weber2025}.

\begin{figure}
\includegraphics[width=\linewidth]{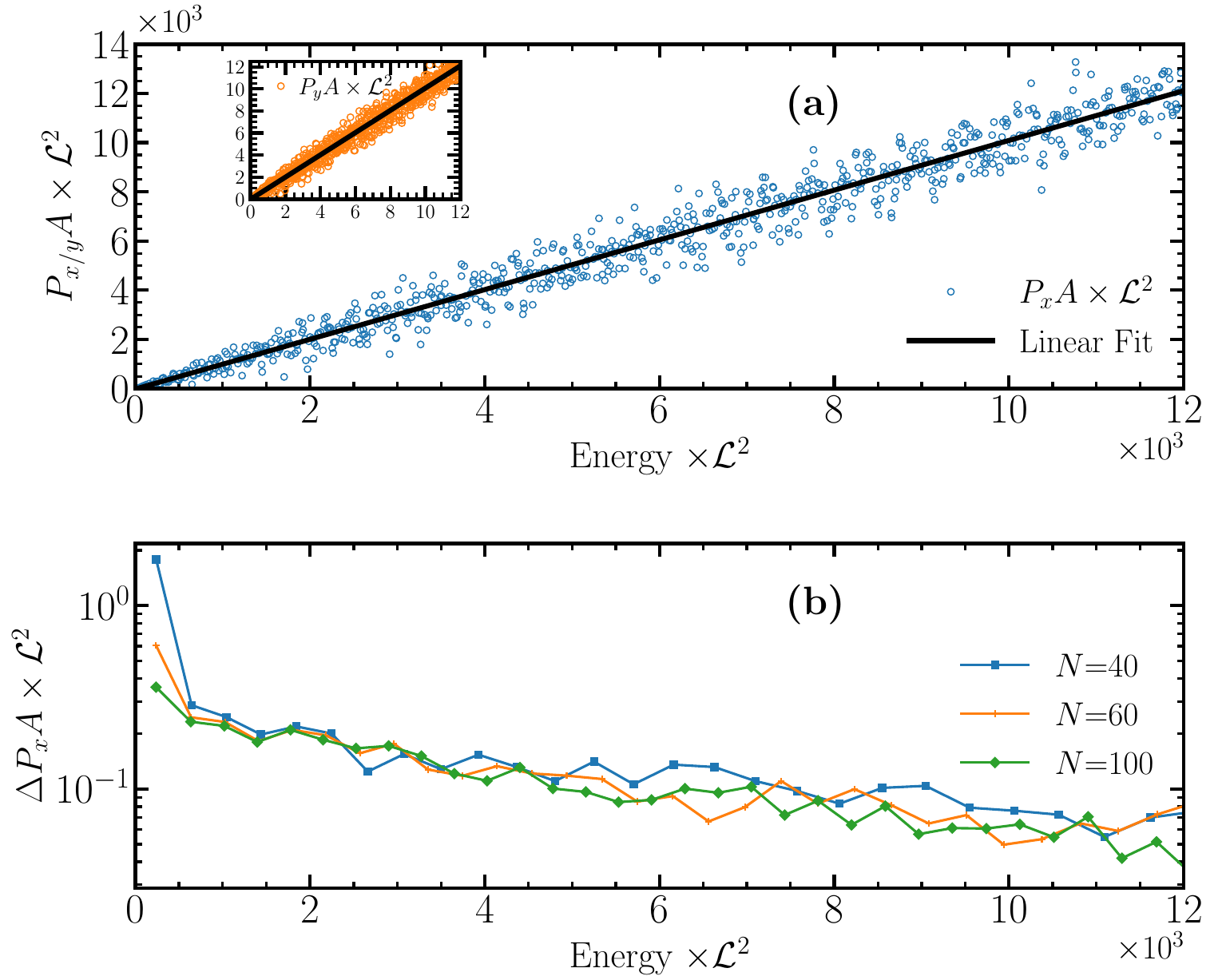}
\caption{
Test of Boyle-Mariotte's law, for a single particle moving freely in the elementary domain of a Sinai billiard, with side lengths $L_x = 1.09\,\mathcal L$, $L_y = 1.00\,\mathcal L$,
quarter circle 
radius $R = 0.5\,\mathcal L$, and resulting volume $A$.
To verify the isotropy
of pressure $P$, the latter is extracted from individual 
eigenenergies of the billiard
via (\ref{HF}), with $\lambda=L_x$ ($L_y$), to obtain $P_x$ 
($P_y$).
Linear fits $f_{P_{x,y}A}(E) = a_{x,y}E + b_{x,y}$ to the products $P_{x,y}A$ for the 800 lowest eigenenergies of the
billiard, to verify Boyle-Mariotte's law in (a) (and the inset),
give $a_x = 1.010$, $b_x\mathcal L^2 = -0.016\times10^3
$, $a_y = 1.008$, $b_y\mathcal L^2 = -0.010\times10^3
$, in very good agreement with
the expected 
$a=1$, $b\mathcal L^2=0$.
(b) Relative fluctuations
$\Delta P_xA = |P_x(E)A - f_{P_xA}(E)|/|f_{P_xA}(E)|$ of $P_xA$, averaged over 25 subsequent eigenstates for each point,
as a function of energy, for three 
grid sizes $N$ of the triangulation employed in the finite element diagonalization routine \cite{brugger2024,wulf2024,kantz2025,weber2025,holm2025}. 
Apart from the low energy range, where 
a sudden decrease of $\Delta P_x A$ with $N$ indicates a lack of numerical convergence for too coarse
a discretization, 
$\Delta P_x A$
smoothly decreases with energy, as to be expected, for decreasing
de Broglie wavelengths. The results in (a) were obtained for $N=100$,
$2N$ 
grid points along the Sinai billiard's quarter-circle, 
a length variation 
$\delta L_{x,y} = 0.01\,\mathcal L$, and third order polynomial interpolation
between the 
resulting eigenvalues, 
to extract the $P_{x,y}$ from the energy levels' derivatives. Analogous results of comparable quality are obtained 
for different ratios $L_x/L_y$ of the billiard's side lengths \cite{kantz2025}.
}
\label{fig-Boyle}
\end{figure}

The resulting values of $P$ are independent of the specific choice of $\lambda$, up to residual fluctuations due to the particle's finite de Broglie wavelength
(tantamount of 
the 
finite size 
of 
$\hbar$), the strength of which will be 
scrutinized in future work. Furthermore, the product of $P$, extracted from the quantized 
particle's parametric energy level dynamics via (\ref{HF}), with the billiard's two-dimensional volume (area) $A$, is given by $PA=E$
(as to be expected for our present, two-dimensional setting \cite{sommerfeld1988,bartelmann2015}), with very good
accuracy,
as 
demonstrated in Fig.~\ref{fig-Boyle}. 
This reproduces \cite{kantz2025} the Boyle-Mariotte law,
again up to fluctuations controlled by the particle's wavelength. Note that the {\em relative} size of these fluctuations decreases with energy, as to be expected 
qualitatively, since the quantum dynamics approaches the limit of ray optics as the particle's momentum increases.

\section{Microscopic Thermal Equilibration and Heat Flux}
\begin{figure}
\includegraphics[width=\linewidth]{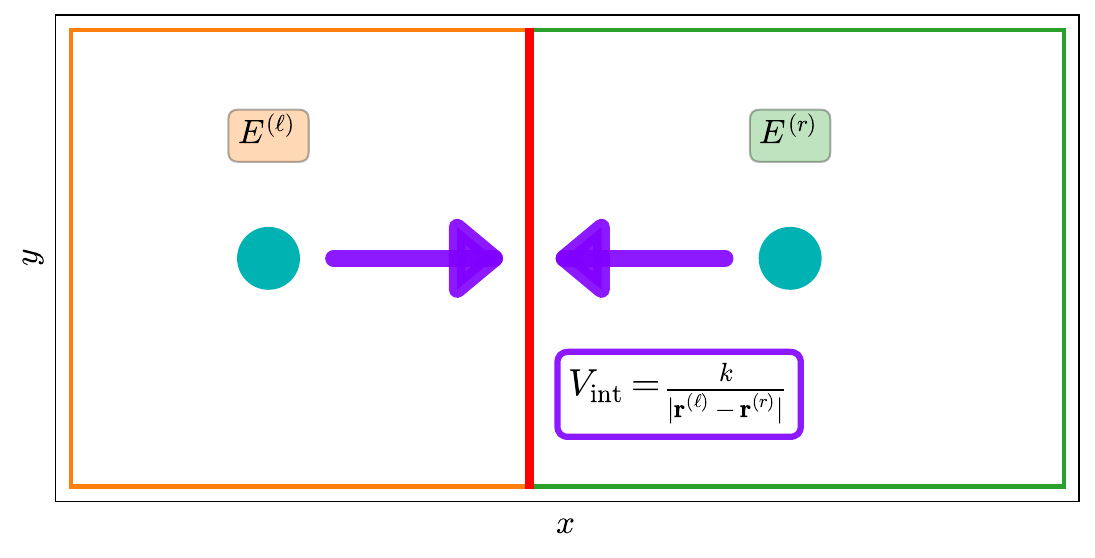}
\caption{
Two 
particles of identical mass, $m\equiv 1$,
placed into two adjacent 2D boxes separated by an immobile wall of finite width $b=0.001\,\mathcal L$, with fixed
side lengths $L_x^{(\ell)} = 1.1\,\mathcal L$, $L_x^{(r)} = 1.3\,\mathcal L$, $L_y=1.4\,\mathcal L$
, where $L_x^{(\ell)}\ne L_x^{(r)}$
to reduce symmetries, and perfectly reflecting boundary conditions are imposed. The particles, initially prepared in a product 
state $\ket{\Psi_0}=\ket{\epsilon^{(\ell )}_{\rm m_0},\epsilon^{({r})}_{\rm n_0}}$ of uncoupled single-particle energy eigenstates
of the left and right box, respectively, exchange energy via the Coulomb interaction $V_\text{int}$,
leading to time dependent local energy expectation values $\langle E^{(\ell ,{r})}(t)\rangle$.
}
\label{sketch-heat}
\end{figure}

While Boyle-Mariotte's law arguably offers a first path to define temperature \cite{kantz2025}, via its statistical mechanics relation to energy \cite{reichl1987,sommerfeld1988,englert2020}, 
we want to introduce temperature as usually done in classical thermodynamics \cite{bartelmann2015}, via the irreversible exchange of energy between 
system constituents, i.e., by the exchange of heat. For this purpose, we extend our billiard model and replace the single particle in a Sinai billiard by 
two particles of which each is confined to one of the two rectangular compartments of one two-dimensional, rectangular domain, see Fig.~\ref{sketch-heat}.
The separating wall between both compartments be immobile (such that no work can be performed) 
and again defines reflecting boundary conditions for both particles. These are further assumed to interact via 
an attractive Coulomb force of strength $k$, leading to a classically mixed phase space, as we verified by running some classical trajectories \cite{holm2025}. 
The corresponding Hamiltonian 
reads, in its quantized form
\begin{eqnarray}
H_{\rm 2p} & = &  h^{(\ell)}+h^{(r)}+V_{\rm int} \nonumber \, , \\
h^{(\ell ,{r})} & = & \frac{({\bf p}^{(\ell ,{r})})^2}{2}+V_{\rm box}^{(\ell ,{r})} \nonumber \, , \\
V_{\rm int} & = & \frac{k}{|{\bf r}^{(\ell)}-{\bf r}^{({r})}|} \, ,
\label{2P}
\end{eqnarray}
with ${\bf r}^{(\ell, {r})}$, ${\bf p}^{(\ell ,{r})}$
the left ($\ell$) and right ($r$) particle's two-dimensional
positions and momenta,
together with the left and right boundary conditions encoded in $V_{\rm box}^{(\ell ,{r})}$, respectively. 

\begin{figure}
\includegraphics[width=\linewidth]{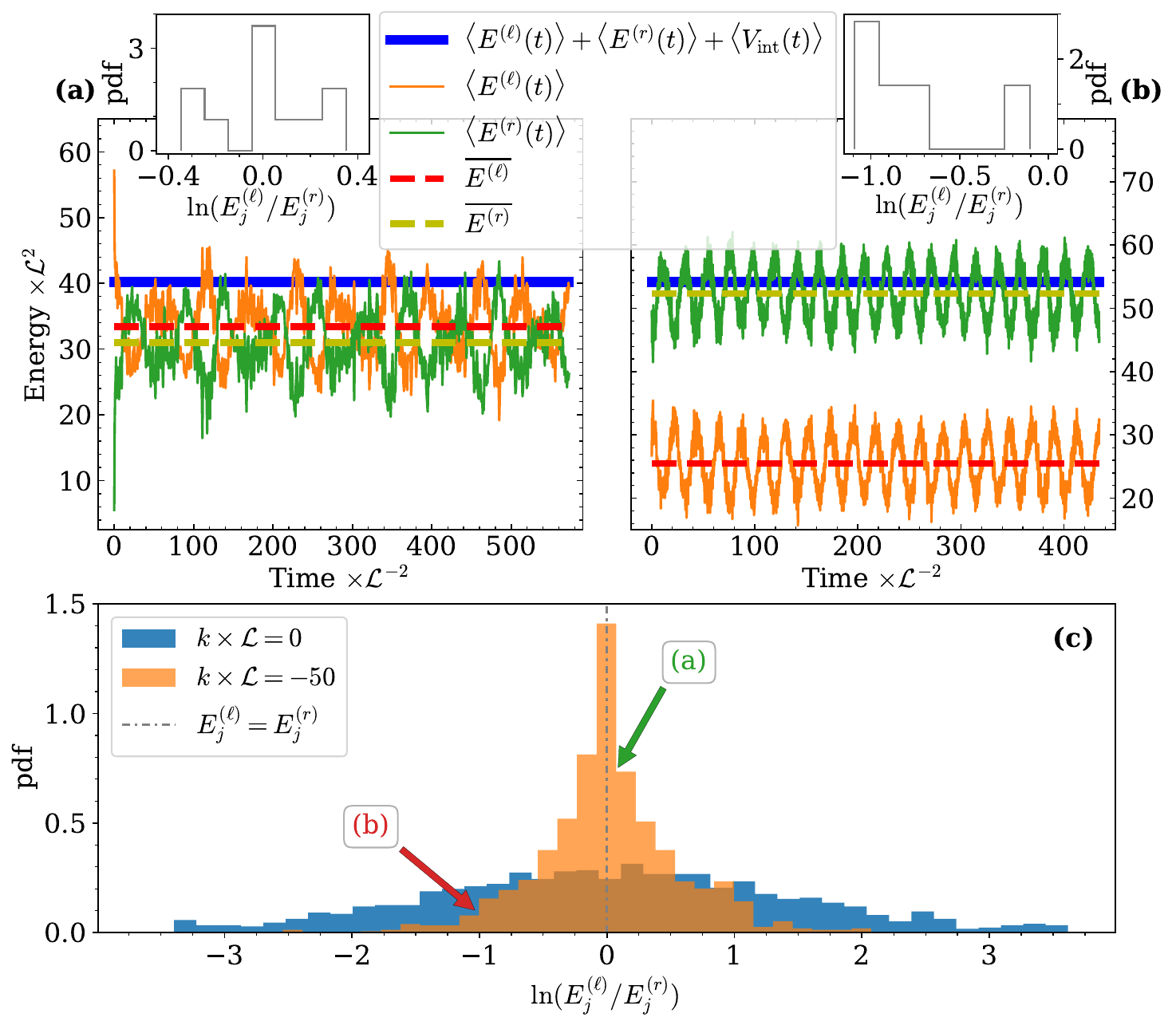}
\caption{Time evolution of the energy expectation values $\langle E^{(\ell ,{r})}(t)\rangle$, for 
initial states 
$\ket{\Psi_0}=\ket{\epsilon^{(\ell )}_{\rm m_0},\epsilon^{({r})}_{\rm n_0}}$ 
with initial local quantum numbers (a) $\rm m_0^{(\ell)} \equiv (2,4)$, $\mathrm{n}_0^{(r)} \equiv (1,1)$, 
(b)  $\rm m_0^{(\ell)} \equiv (1,3)$, $\mathrm{n}_0^{(r)} \equiv (4,1)$,  
together with the (conserved) total energy obtained as the sum of $\langle E^{(\ell, {r})}(t)\rangle$ and $\langle V_\text{int}(t)\rangle$, for $k\mathcal L=-50$.
(c) Distribution of the energy balance ratios $\ln(E_j^{(\ell)} / E_j^{(r)})$ for the first (energy-ordered) 1000 eigenstates, for vanishing interaction strength $k\mathcal L=0$, and for $k\mathcal L=-50$.
The insets in (a,b) show the equivalent distributions which only include those (eleven (a) and five (b)) eigenstates $\ket{E_j}$ with which the respective initial state $\ket{\Psi_0}$ has an overlap 
$|\langle\Psi_0|E_j\rangle|^2 \geq 2\%$. The arrows in (c), in turn, indicate the energy balance ratios of those single 
$\ket{E_j}$ which have the largest overlap $|\langle\Psi_0|E_j\rangle|$
with the initial states evolved in (a,b). The system side lengths are, as defined before, $L_x^{(\ell)} = 1.1\,\mathcal L$, $L_x^{(r)} = 1.3\,\mathcal L$, $L_y=1.4\,\mathcal L$.
The energy scales here covered are small compared to the ones in Fig.~\ref{fig-Boyle}, since the numerically accessible energy range for two particles is
considerably smaller than that for one particle.
}
\label{equi}
\end{figure}

We now prepare each of the particles in an energy eigenstate of the uncoupled, local single-particle Hamiltonian $h^{(\ell,{r})}$, 
at (eigen) energies $\epsilon^{(\ell )}_{\rm m_0}> \epsilon^{({r})}_{\rm n_0}$ 
or $\epsilon^{(\ell )}_{\rm m_0}< \epsilon^{({r})}_{\rm n_0}$ \footnote{The multi-indices $\rm m_0,n_0$ are each tuples of positive
integer quantum numbers which indicate the respective eigenstate's excitations in the $x$- and $y$-direction, respectively.},
let the corresponding two-particle initial state 
$\ket{\Psi_0}=\ket{\epsilon^{(\ell )}_{\rm m_0},\epsilon^{({r})}_{\rm n_0}}$
evolve under the action of the unitary $U(t)$ generated by the {\em coupled} Hamiltonian 
$H_{\rm 2p}$, and monitor the energy expectation values 
$\langle E^{(\ell ,{r})}(t)\rangle =  \bra{\Psi_0} U(t)^\dagger h^{(\ell ,{r})} U(t) \ket{\Psi_0}$
of the left and of the right particle as a function of time, with a 
typical result displayed in Fig.~\ref{equi} (a).
As we see, energy is irreversibly redistributed between both particles on rather short time scales, and with residual fluctuations of an amplitude which is 
strictly smaller
than the energy loss/gain of the left/right particle during the initial transient phase of the evolution. 
As a counter example, in Fig.~\ref{equi} (b), we 
show the time evolution seeded by an initial condition which we associate with the regular domains of classical phase
space \cite{holm2025}: The time evolution does not exhibit equilibration, and the observed signals' frequency content is scarce (consistent with the association 
with classically regular motion).

Upon averaging over the left and right particle's energy expectation values' residual fluctuations (a) or oscillations (b), 
the asymptotic energies of both
particles 
can be identified with the (time independent) diagonal approximations $\overline{E^{(\ell ,{r})}}$ of their energy expectation values 
$\langle E^{(\ell ,{r})}(t)\rangle $.
Given the spectral decomposition of $U(t)$ into the projectors onto the coupled two-particle energy eigenstates $\ket{E_j}$
of 
$H_{\rm 2p}$, these read \footnote{For fixed parameters of $H_{\rm 2p}$, only the
	weights $|\langle\Psi_0|E_j\rangle|^2$ need to be evaluated anew when changing the initial local energies which define 
	$\ket{\Psi_0}=\ket{\epsilon^{(\ell )}_{\rm m_0},\epsilon^{({r})}_{\rm n_0}}$.}
\begin{equation}
\overline{E^{(\ell, {r})}} = \sum_j |\langle\Psi_0|E_j\rangle |^2 \sum_{m,n} |\langle\epsilon^{(\ell )}_{\rm m},\epsilon^{({r})}_{\rm n}|E_j\rangle |^2 \epsilon^{(\ell, r)}_{\rm m, n}.
\end{equation}

Note, however, that the dynamical emergence of equilibration, for 
unstable, or its absence, for 
stable initial conditions, is already imprinted 
into the coupled eigenstates which contribute to the dynamics via non-vanishing $|\langle\Psi_0|E_j\rangle |^2$, as illustrated in the insets 
of Fig.~\ref{equi} (a,b), and in (c): The (logarithmic) energy balance 
ratio $\ln(E_j^{(\ell)}/E_j^{({r})})$ is strongly biased against zero when the two-particle system is launched in a regular phase space domain (b), while it concentrates symmetrically around zero for eigenstates contributing to dynamics seeded by an unstable 
initial condition (a). The latter feature, manifest 
on the level of the propagation 
of specific initial conditions, is likewise encoded in all eigenstates, as illustrated in Fig.~\ref{equi} (c), for the lowest lying 1000 coupled eigenstates of $H_{\rm 2p}$: 
Their
logarithmic energy balance ratio distribution
exhibits a 
pronounced concentration around zero, in stark contrast to the distribution extracted from the 
uncoupled ($k=0$) eigenstates.
The ($k\ne 0$) eigenstate with dominant contribution to the dynamics depicted in (a) 
is located in the centre of this distribution, while the eigenstate principally contributing to the dynamics of (b) sits in the distribution's wings, 
as indicated by the (red and green) arrows in (c).

 Given the evidence for equilibration provided in 
 Fig.~\ref{equi} (a,c), 
we can follow the standard procedure to define 
heat, temperature and energy \cite{sommerfeld1988,kittel1984,reichl1987,bartelmann2015,englert2020}. Heat is 
given by the net energy 
exchange $\delta Q = \epsilon^{(\ell)}_{\rm m_0}-\overline{E^{(\ell)}}=  \overline{E^{({r})}}-\epsilon^{({r})}_{\rm n_0} $, on time 
scales longer than the transient following the switching, at $t=0$,
of the particle-particle interaction. It can 
be written
as
\begin{eqnarray}\label{eq:dQ}
	\delta Q & = & \sum_j\Delta \rho_{jj}^{(\ell)}\epsilon^{(\ell)}_{\rm j} = {\rm tr}^{(\ell)}(\Delta \rho^{(\ell)} h^{(\ell)})\, , \nonumber
\end{eqnarray}
where $\Delta \rho^{(\ell)} = \rho^{(\ell)}(t=0) - \overline{\rho^{(\ell,{r})}}$, with the reduced single-particle state $\rho^{(\ell)}$ represented in the
left particle's uncoupled eigenstates (which are the 
individual uncoupled particles' natural microstates). The overline again indicates
the diagonal 
approximation, which 
characterizes 
the 
new equilibrium state. The latter maximizes the number of those microstates of 
both particles
which are compatible with the fixed two-dimensional volumina
$A^{(\ell,{r})}$, as well as with
the conserved total energy
$E_\text{tot}=\epsilon^{(\ell)}_{\rm m_0} + \epsilon^{({r})}_{\rm n_0} + \langle V_\text{int} \rangle (t=0)$ (note that, since $\langle V_\text{int}\rangle \approx \text{const}$, also $E = \epsilon^{(\ell)}_{\rm m_0} + \epsilon^{({r})}_{\rm n_0}$ is -- on average -- conserved). 
This maximization
enforces the equality of the derivatives of the entropies of the individual particles on the left and on the right, with respect to their local energies, evaluated 
at their local equilibrium energies 
$\overline{E^{(\ell ,{r})}}$ \cite{sommerfeld1988,reichl1987,bartelmann2015}.

Given 
the interacting two-particle spectrum and the associated interacting eigenstates, 
the local entropies are given as
\begin{equation}\label{}
S^{(\ell,{r})}=-\sum_j\rho_{jj}^{(\ell,{r})}\ln\rho_{jj}^{(\ell,{r})}\, ,
\label{loc-ent}
\end{equation}
where the
$\rho_{jj}^{(\ell,{r})}$ are
obtained
upon trace 
of 
$U\ket{\Psi_0}\bra{\Psi_0}U^\dagger$ 
over the 
right/left particle, again represented in the left/right particle's 
uncoupled eigenstates.
Sampling over different instances of $E=\epsilon^{(\ell )}_{\rm m_0}+\epsilon^{({r})}_{\rm n_0}$ allows to obtain the values of the local entropies at different 
emergent equilibrium subsystem energies $\overline{E^{(\ell,r)}}$, which can be fitted to a logarithmic dependence of $S^{(\ell,r)}$ on $\overline{E^{(\ell,r)}}$
(this functional dependence being implied by 
the initially quoted principle of equal a priori probabilities).
The local inverse  temperatures $1/T^{(\ell, {r})}$ then follow as the energy derivatives of $S^{(\ell,r)}(\overline{E^{(\ell,r)}})$.

\begin{figure}
\includegraphics[width=\linewidth]{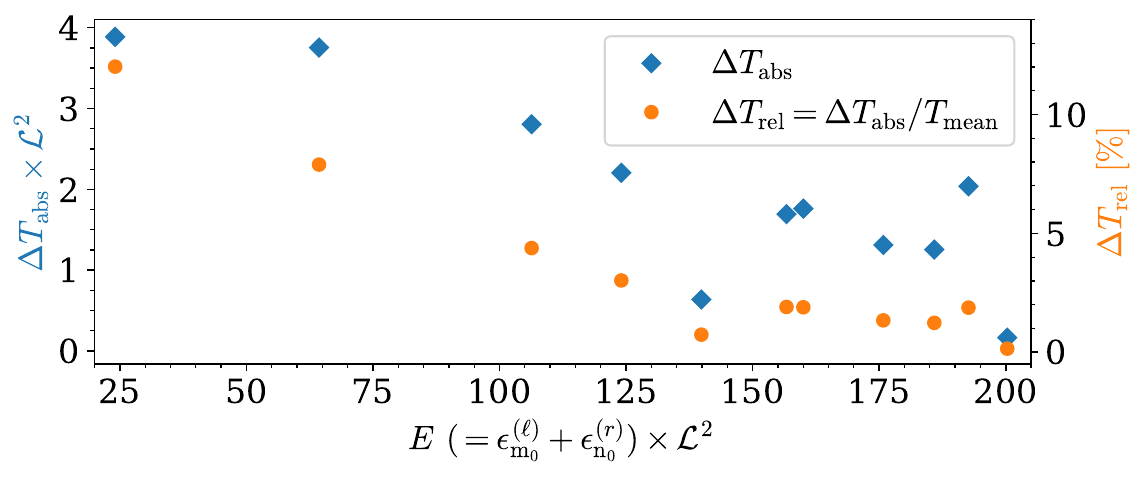}
\caption{Absolute, $\Delta T_\text{abs}$, and relative, 
$\Delta T_\text{rel}$, offsets between the right and the left particle's equilibrium temperatures  $T^{(\ell, r)}$ (with $k_B \equiv 1$), as inferred from the 
energy derivatives of $S^{(\ell,{r})}$, 
for initial states $\ket{\Psi_0}=\ket{\epsilon^{(\ell )}_{\rm m_0},\epsilon^{({r})}_{\rm n_0}}$ selected 
by two criteria: (i) Large initial energy 
mismatch,
i.e.~$\epsilon_{\text{m}_0}^{(\ell)} \gg \epsilon_{\text{n}_0}^{(r)}$ (or vice versa),
and (ii) small final energy 
mismatch, i.e.~$\overline{E^{(\ell)}} \approx \overline{E^{(r)}}$.
$\Delta T_\text{abs}$ and $\Delta T_\text{rel}$ initially clearly decrease with increasing energy
$E = \epsilon_{\text{m}_0}^{(\ell)} +  \epsilon_{\text{n}_0}^{(r)}$, as expected
as one approaches 
the short wavelength limit. The 
saturation for $E\mathcal L^2\geq 150$ is attributed to the finite numerical error of the eigenenergies \cite{holm2025}.
}
\label{temp}
\end{figure}

For the equilibrium energies $\overline{E^{(\ell ,{r})}}$ extracted from dynamics as depicted in Fig.~\ref{equi} (a)
to be consistent with
the standard definition of thermal equilibrium, they 
have to 
correspond to equal temperatures $T^{(\ell)}=T^{(r)}$ 
inferred from $S^{(\ell,{r})}$ \cite{sommerfeld1988,reichl1987,bartelmann2015}. 
This is confirmed by 
Fig.~\ref {temp},
within a finite error margin, 
likely due to the involved particles' finite wavelengths, at finite excitation energies. 
The figure shows that the residual absolute temperature offset $\Delta T_\mathrm{abs} = |T^{(\ell)} - T^{(r)}|$, as well as the relative temperature offset 
$\Delta T_\mathrm{rel} = \Delta T_\mathrm{abs}/T_\mathrm{mean}$, with $T_\mathrm{mean} = (T^{(\ell)}+T^{(r)})/2$,
indeed systematically decrease with increasing 
energy $E=\epsilon^{(\ell)}_\mathrm{m_0} + \epsilon^{(r)}_\mathrm{n_0}$. This decrease, however, saturates for $E\mathcal L^2 \geq 150$, which we attribute to the worsening of numerical convergence for higher eigenenergies.

\section{Conclusion}
For the textbook-like example here considered (with the 
interacting particles' excitation spectrum as non-trivial input), we achieve
a 
consistent definition of the equilibrium state variables $P$, $T$ and $S$, which secures the first law.
Since lifting the thermal isolation between the left and right particle, by switching on 
their interaction term in (\ref{2P}), 
corresponds to relaxing one 
constraint, the number of the two-particle system's
microstates increases during the
equilibration process, which 
implies the second law. The interaction energy itself exhibits 
small fluctuations during equilibration, 
and becomes negligible for increasing total energy, since,
on average, it doesn't 
increase with 
energy. Work can be defined as in equilibrium thermodynamics, by 
$\delta W=-PdA$, with $P$ given by (\ref{HF}), directly from
single-particle energy level velocities, and heat as $\delta Q=TdS$,
with the energy dependence 
of $S$ as defined in (\ref{loc-ent}) (via the energy dependence of the $\rho_{jj}^{(\ell,{r})}$) the fundamental quantity.

\section*{Acknowledgements}
We are indebted 
to Jonathan Brugger for developing the original version of the 
finite element code here employed to generate our numerical results, to Gabriel Dufour (G.D.) for helpful discussions at an early stage of this project, and to 
G.D., Cord A.~M\"uller and Beno\^{\i}t Zumer for critical comments on the manuscript. M.B. thanks the Georg H.~Endress Stiftung for funding and support.
 
\bibliographystyle{apsrev4-2} 
\bibliography{thermo-Holm}

\end{document}